\documentstyle[twocolumn,epsf,seceq]{jpsj}

\def\nle{\ \raise.3ex\hbox{$<$}\kern-0.8em\lower.7ex\hbox{$\sim$}\ }
\def\nge{\ \raise.3ex\hbox{$>$}\kern-0.8em\lower.7ex\hbox{$\sim$}\ }
\def\alphat{\alpha (\tau)}

\def\chiomegat{\chi''(\omega; t)}
\def\chiomegatw{\chi''(\omega; t_{\rm w})}

\def\DelCone{1 - C(\tau; t_{\rm w})}
\def\DelC{\Delta C(\tau; t_{\rm w})}
\def\qea{q_{\rm EA}}
\def\Tc{T_{\rm c}}
\def\tw{t_{\rm w}}
\def\Upeff{\Upsilon^{\rm eff}}

\def\runtitle{Numerical Study on Aging Dynamics in  the 3D Ising
  Spin-Glass Model. II.}
\def\runauthor{Tatsuo {\sc Komori}, Hajime {\sc Yoshino} and Hajime 
{\sc Takayama}}

\title{Numerical Study on Aging Dynamics in  the 3D Ising Spin-Glass
   Model. II. Quasi-Equilibrium Regime of Spin Auto-Correlation
   Function\footnote{To appear in J. Phys. Soc. Jpn. {\bf 69} (2000) No. 4}}
\author{Tatsuo {\sc Komori}\footnote{Present address: Hydrographic
      Department, Maritime Safety Agency, 5-3-1 Tsukiji, Chuo-ku,
      Tokyo 104-0045},
 Hajime {\sc Yoshino}\footnote{E-mail: yhajime@ginnan.issp.u-tokyo.ac.jp} and
 Hajime {\sc Takayama}\footnote{E-mail: takayama@issp.u-tokyo.ac.jp}}
\inst{Institute for Solid State Physics, the University of Tokyo,
7-22-1 Roppongi, Minato-ku, Tokyo 106-8666}
\recdate{\today}
\abst{Using Monte Carlo simulations, we have studied isothermal aging
of three-dimensional Ising spin-glass model focusing on 
quasi-equilibrium behavior of the spin auto-correlation function. Weak 
violation of the time translational invariance in the
quasi-equilibrium regime is analyzed in terms of {\it effective
  stiffness} for droplet excitations in the presence of domain walls.
Within the range of computational time window, we have confirmed that
the effective stiffness follows the expected scaling behavior with
respect to the characteristic length scales associated with droplet
excitations and domain walls, whose growth law has been extracted from 
our simulated data.
Implication of the results are discussed in relation to experimental 
works on ac susceptibilities.}

\kword{spin glass, slow dynamics, aging, droplet theory}

\begin{document}
\sloppy
\maketitle

\section{Introduction}

In recent years aging dynamics in spin glasses has been extensively 
studied.~\cite{BCKM,VincHO,Nordblad,Weissman,Rieger-95} A basic
experimental protocol is isothermal aging which is relaxation after
a spin-glass (SG) system is quenched from a high temperature above the
SG temperature $T_{\rm c}$ to a low temperature below 
$T_{\rm c}$. Experimental studies of the relaxation of dc and ac
susceptibilities during isothermal aging have revealed remarkable
waiting time effects. The experiments have stimulated much theoretical
interests and many theoretical ideas have been proposed most of which
could account for qualitative features of the aging 
effects.~\cite{BCKM,FH-88-NE,KoperH-88,Sibani-H}
Real tests on the theories are possible by investigating quantitative 
features of the aging effects, especially scaling properties.
Progress in this direction, which may shed light on the controversial
issues on nature of the SG phase, is still left to be done to
a large extent.

The droplet theory,~\cite{FH-88-NE,FH-86-PRL,BM-87-Heidel,FH-88-EQ}
particularly the one proposed by Fisher and Huse~\cite{FH-88-NE} 
contains concrete ansatz on possible scaling laws of the aging effects 
which are well amenable to be tested quantitatively by experiments and 
numerical simulations.
In our previous paper  (hereafter referred to as I),\cite{oursI} 
we have studied the relaxation of energy of the three-dimensional (3D)
Gaussian SG model with nearest neighbor interactions below 
$\Tc$ using Monte Carlo (MC) simulations. We analyzed the data comparing
with the droplet theory. A key quantity in the droplet theory is 
the typical separation between domain walls $R(t)$ which grows
irreversibly by isothermal aging. 
Most of the scaling laws involve this characteristic length.
Indeed in I we confirmed that the relaxation of excessive energy per
spin $\delta e_{T}(t)$ at time $t$ with respect to the equilibrium
value follows the expected scaling law, 
\begin{equation}
\delta e_{T}(t) 
\sim \tilde{\Upsilon}(R(t)/L_0)^\theta / (R(t)/L_0)^d, 
\label{eq:ene-sim}
\end{equation} 
when we used $R(t)$ obtained by our independent simulation (see
below). In the above equation $\tilde{\Upsilon}$ and $L_{0}$ are
characteristic energy and length scales respectively and $d$ is the
dimension of the space which is $3$ here. The energy exponent obtained
is $\theta = 0.20 \pm 0.03$ which agrees with the result of the defect
energy analysis at $T=0$.~\cite{BM-84-theta} 

The simulated $R(t)$ mentioned above was obtained by analyzing
the replica-overlap function and found, in agreement with the previous 
works~\cite{Kisker-96,Marinari-98-VFDT} (but see also \cite{Huse-91}),
that the growth law of $R(t)$ is well approximated by the form,
\begin{equation}
   R(t) \sim L_0(t/\tau_0)^{1/z(T)}, 
\label{eq:Rt-sim}
\end{equation} 
where $\tau_0$ is a certain characteristic unit of time. More
explicitly, our data of $R(t)$ at each $T$ are well fitted to a 
function $bt^{1/z}$ with adjustable parameters $b$ and $z$. The
latter depends on $T$ and is expressed as 
\begin{equation}
1/z(T) \simeq 0.17T,
\label{eq:z_T}
\end{equation}
at  $T <0.7$. Here and in the following, the Boltzmann constant
$k_{\rm B}$ is set to $1$, and temperature $T$ is measured by the 
unit of variance of the distribution of the interaction bonds 
$J$ which is set to $1$. As Kisker et al~\cite{Kisker-96} already 
pointed out, however, the simulated $R(t)$ can be fitted also to a 
logarithmic function $R_0 + b ({\rm ln}t)^{1/\psi}$ with adjustable 
parameters $R_0, b$ and $\psi$, where $\psi$ is independent of
temperature. This form, in the limit $R(t) \gg R_0$, is 
compatible with the asymptotic form proposed by the droplet theory, 
\begin{equation}
   R(t) \sim L_0[(T/\Delta){\rm ln}(t/\tau_0)]^{1/\psi}, 
\label{eq:Rt-FH}
\end{equation}
where $\Delta$ is a characteristic unit for energy barriers 
associated with thermal activation of droplet excitations. $L_0$ and 
$\tau_0$ in eqs.(\ref{eq:Rt-sim}) and (\ref{eq:Rt-FH}) are different 
sets of characteristic length and time scales. 

The above-mentioned results may be summarized as
\begin{enumerate}
\renewcommand{\labelenumi}{\arabic{enumi})}
\item Various quantities (such as $\delta e_{T}(t)$) simulated the
  isothermal aging and the characteristic length scales (such as
  $R(t)$) obtained by the independent simulation satisfy the scaling
  forms (such as eq.(\ref{eq:ene-sim})) which are predicted by the
  droplet theory.
\item The growth law of $R(t)$ is well fitted to the power-law form
  eq.(\ref{eq:Rt-sim}), but it can be also fitted to a certain 
  logarithmic form which is compatible with the logarithmic growth law 
  eq.(\ref{eq:Rt-FH}).
\end{enumerate}
Feature 2) suggests that the computational time window available by
modern computers is not enough large for us to determine definitely a
true asymptotic behavior of $R(t)$ in the limit $t \rightarrow
\infty$.  Therefore we call the above results seen in the present 
computational time window the {\it pre-asymptotic} behavior of aging.
A similar situation was already encountered in the pioneering
numerical work by Huse~\cite{Huse-91} who could confirm the density of
domain walls follows the expected scaling form
$1/L_{0}^{d} (R(t)/L_{0})^{d_{\rm s}-d}$  with reasonable value of the 
fractal dimension $d_{\rm s}$ but with the growth law being in some 
pre-asymptotic form.

In the present paper,~\cite{Komori} we study isothermal aging behavior 
of the spin auto-correlation function in the so-called
quasi-equilibrium regime. Here we focus our study on its scaling
properties, or more explicitly, whether the simulated correlation
function can be described by a certain scaling form which is derived
from the droplet theory, i.e., feature 1) above. To this end, we have
performed extensive
MC simulations on the 3D Gaussian Ising SG model in
the same time window as that in our previous work I. 

The spin auto-correlation function is defined as,
\begin{equation}
        C(\tau; \tw) = \overline{ C_i(\tau; \tw)}, 
\label{eq:def-Cor}
\end{equation}
with
\begin{equation}
        C_i(\tau; \tw) = \langle S_i(\tau+\tw)S_i(\tw)\rangle,
\label{eq:def-iCor}
\end{equation}
where $S_{i}(t)$ is the sign of the Ising spin at site $i$ at time
$t$. At time $t=0$, the system is prepared in a random initial
configuration and we let it relax at $T$ below $\Tc$ using
conventional heat-bath MC dynamics afterwards.
Here and in the following time $t$ is measured by a MC 
step (MCS). By $1$ MCS, all Ising spins are updated once.
The auto-correlation function over some time intervals $\tau$ at various 
waiting times $\tw$ are observed during a MC run of isothermal aging.
The over-line in eq.(\ref{eq:def-Cor}) denotes the averages 
over sites and different realizations of interactions (samples), and 
the bracket in eq.(\ref{eq:def-iCor}) the average over thermal noises
(or different MC runs). 

The quasi-equilibrium regime, on which we focus in this study, is the 
time regime such that the time separation $\tau$ is relatively small 
compared with the waiting time $\tw$.  
Although behavior of the correlation function is not extremely
different from the equilibrium limit within this regime, it exhibits
systematic deviations from the ideal equilibrium behavior, namely,
clear waiting time dependence, i.e., weak violation of time
translational invariance. It is directly comparable with the
measurement of relaxation of the Fourier spectrum of spontaneous
magnetic fluctuations \cite{AlbaHOR-87,RefregierOHV-88} in
isothermal aging. Note that the Fourier component at a certain
frequency $\omega$ is well defined only at large enough waiting times 
such that $\omega\tw \gg 1$. 
For the same reason, the ac susceptibility is also considered 
to measure mainly relaxation of response of the system in this regime.
From the point of view of the droplet theory, the quasi-equilibrium 
regime is defined more precisely as the time regime such that the 
typical size $L(\tau)$ of droplet excitations which take place in the 
time scale of $\tau$ is much smaller than the typical separation 
$R(\tw)$ of domain walls present after waiting time $\tw$.
On the other hand, let us recall for clarity that the so-called aging (or 
off-equilibrium) regime is the time regime such that $\tau$ is as large as 
or much larger than the  waiting time $\tw$. 
More precisely, the aging regime is defined as the time regime where 
$R(t=\tau+\tw)$ is larger than $R(\tw)$.
The relaxation of the dc 
susceptibilities, which include the thermoremanent magnetization
(TRM) and zero-field cooled susceptibilities (ZFC), can reflect strong 
non-stationarity within the aging regime. 

The droplet theory introduced a phenomenological concept called 
{\it effective stiffness} which characterizes reduction of the 
excitation gap of small scale droplet excitations due to the presence
of domain walls.  The latter is considered to explain the waiting time 
effect within the quasi-equilibrium regime.
We directly compare our results with this scaling ansatz.
As far as we know, no direct quantitative analysis has been 
performed to test the scaling ansatz except for an experimental 
study of relaxation of the ac susceptibility in a two-dimensional 
SG system.~\cite{Schins-93} 
The previous numerical simulations have analyzed the non-stationarity
of the spin auto-correlation within the aging 
regime,\cite{Rieger-95,Rieger-93,Kisker-96} but the analysis of the
weak non-stationarity within the quasi-equilibrium regime has not been 
carried out.

We found that the spin auto-correlation function $C(\tau; \tw)$ in the
quasi-equilibrium regime is consistently explained by the droplet 
theory in the following senses.  
i) As a function of $L(\tau)/R(\tw)$, $L(\tau)$ being the 
characteristic length of droplet excitations introduced above, 
it obeys a scaling law which is consistent with the scenario in terms 
of the effective stiffness.~\cite{FH-88-NE}  
ii) Result i) is obtained when $L(\tau)$ and $R(\tw)$ are assumed to
obey the same growth law. The latter is a fundamental assumption within
the droplet theory for spin glasses, while in a simple ferromagnet 
domain growth and droplet excitations are completely different kinds 
of processes. 
iii) We have confirmed result i) explicitly by making use of the growth 
law of $R(t)$, which is obtained by our simulation and is approximated 
by the power law of eq.(\ref{eq:Rt-sim}). Correspondingly, we have used  
\begin{equation}
L(\tau) \sim L_0 (\tau/\tau_0)^{1/z(T)},
\label{eq:Ltau-sim}
\end{equation}
for $L(\tau)$. These results add a further support for feature 1)
of the aging dynamics mentioned before, though they don't
resolve the ambiguity pointed out as feature 2). We will argue that
a similar situation to feature 2) is also encountered in the
experimental studies up to now.   

The present paper is organized as follows. In the next section we 
briefly review the scaling ansatz of the droplet theory.
In \S 3 we present our simulated data on the spin correlation and
response functions in isothermal aging. In particular, behavior of
$C(\tau; \tw)$ in the quasi-equilibrium regime ($\tau \ll \tw$) is
examined in detail and is interpreted by the droplet theory.
We discuss implications of the present results with some other
numerical and experimental results in \S 4. 

\section{Droplet Picture}

Here we  review the droplet theory ~\cite{FH-88-NE,FH-88-EQ} by 
Fisher-Huse (FH) concentrating on the influence of domain walls on 
the droplet excitations during isothermal aging.
Let us begin by recalling the ideal equilibrium situation, where 
there should be no non-stationarity and the behavior of the
auto-correlation function defined in eq.(\ref{eq:def-Cor}) becomes
only a function of the time difference $\tau$.

Equilibrium thermal fluctuations are considered as droplet 
excitations~\cite{FH-88-EQ} from a ground state, each of which
is a global flip of a droplet (cluster) of spins within
a distance $L/2$ from a certain given site $i$. 
The latter can be considered as a simple two-state system with 
a free-energy excitation gap $F_{L}(i)$ and a thermal activation time 
$\tau_{L}(i)$. The typical value $F^{\rm typ}_L$ of the gap $F_{L}(i)$
scales as, 
\begin{equation}
   F^{\rm typ} _L \sim  \Upsilon (L/L_0)^\theta,
\label{eq:FL-FH}
\end{equation}
where $\Upsilon$ is the stiffness constant of domain wall on the
boundary of the droplet. The thermal activation time $\tau_{L}(i)$
is related to the energy barrier of the droplet $B_{L}(i)$ as,
\begin{equation}
B_{L}(i)=T \ln[\tau_{L}(i)/\tau_0].
\end{equation}
The typical energy barrier $B^{\rm typ}_L$ of the energy barrier
scales as,
\begin{equation}
B^{\rm typ}_L \sim \Delta (L/L_0)^\psi,
\label{eq:BL-FH}
\end{equation}
where exponent $\psi$ satisfies $\theta \le \psi \le d-1$.

The probability distribution function $\rho_{L}(F)$ 
is assumed to follow a one-parameter scaling form,
\begin{equation}
       \rho_{L}(F) = {1 \over F^{\rm typ}_{L}}
       {\tilde \rho}\left( {F \over  F^{\rm typ}_{L}} \right),  
\label{eq:rho-FH}
\end{equation}
where the typical gap $F^{\rm typ}_{L}$ is given by
eq.(\ref{eq:FL-FH}). A very important property of the scaling function
${\tilde \rho}(x)$ is that it has finite intensity at $x=0$, 
${\tilde \rho}(0)>0$, which allows droplet excitations even at very
low temperatures. The latter is considered to be responsible for many 
non-trivial properties of the SG phase in 
equilibrium.\cite{FH-88-EQ} 
It is also the case for dynamics in the quasi-equilibrium regime of 
the present interest.
A similar one-parameter scaling form for $\Psi_L(B)$ is assumed as,
\begin{equation}
       \Psi_{L}(B) = {1 \over B^{\rm typ}_L}
       {\tilde \Psi}\left( {B \over B^{\rm typ}_L } \right), 
\label{eq:Psi-FH}
\end{equation}
where $B^{\rm typ}_L$ is the typical energy barrier of 
eq.(\ref{eq:BL-FH}) and ${\tilde \Psi}(x)$ is a scaling function.

The droplet excitations at different scales $L$ which differ by more 
than factor $2$ are considered as approximately independent from each
other. Then using the two-state model, the equilibrium behavior of the
spin auto-correlation function is obtained as,
\begin{equation}
C(\tau) \equiv  \lim_{\tw \rightarrow \infty}
\overline{C_{i}(\tau;\tw) }
\sim  \overline{ \prod^\wedge_{L; \tau_{L}(i)<\tau} 
\tanh^2 \left({F_{L}(i) \over 2T} \right) },
\label{eq:cal-Cor-1}
\end{equation}
where ${\hat \Pi}_{L}\ldots$ represents multiplicative contributions
of the relaxations of two-state systems at various scales 
$L=2^{n}L_{0}$ with $n=0,1,2,\ldots\infty$ enclosing the site $i$. 
The factor $\tanh^2 \left({F_{L}(i)/ 2T} \right)$ is the
equilibrium overlap of the two-state system of energy gap $F_{L}(i)$
with respect to the lower state (ground state).
The constraint $\tau_{L}(i) < \tau$ is due to the fact that the
droplets whose relaxation time is shorter than $\tau$ can relax
within the period of $\tau$ while others with $\tau_{L}(i) > \tau$
cannot relax appreciably and just give multiplicative contributions of
$1$. At low enough temperatures, the product in
eq.(\ref{eq:cal-Cor-1}) can be expanded into a sum and the averages
over sites (and equivalently over samples) can be performed using the
probability distribution functions of the free-energy gaps and
free-energy barriers. One obtains, 
\begin{eqnarray}
  1 & - & C(\tau) \nonumber \\
  & \sim &  \int_{L_0}^\infty {{\rm d}L \over L} 
     \int_0^{T{\rm ln}(\tau/\tau_0)}
  {\rm d}B\Psi_{L}(B) \int {\rm d}F\rho_{L}(F){\rm
    e}^{-F/T} \nonumber \\
    & \sim & \int_{L_0}^\infty {{\rm d}L \over L} \Phi
    \left(\frac{L}{L(\tau)} \right) 
   \int {\rm d}F\rho_{L}(F){\rm e}^{-F/T}, \nonumber \\
   & \sim &  T  \int_{L_0}^\infty {{\rm d}L \over L} \Phi
   \left(\frac{L}{L(\tau)} \right) \rho_{L}(0), 
\label{eq:cal-Cor-2}
\end{eqnarray}
where we introduced,
\begin{equation}
\Phi(x)=\int_{0}^{1/x^\psi} {\rm d}y \tilde{\Psi} (y),
\label{eq:filter}
\end{equation}
and the characteristic length $L(\tau)$ defined as,
\begin{equation}
L(\tau)=L_{0}[(T/\Delta)\ln (\tau/\tau_{0})]^{1/\psi}.
\label{eq:L_tau-FH}
\end{equation}
The function $\Phi(x)$ satisfies $\Phi(0)=1$ due to the normalization
of the distribution function $\Psi(B)$ and is expected to remain close
to $1$ for  $x < 1$ and decays rapidly for $x \gg 1$.
The characteristic length $L(\tau)$ is understood as a typical size of 
droplet which can flip in the period of $\tau$. In the last equation of 
eq.(\ref{eq:cal-Cor-2}), we assumed that temperature is low enough
so that only the term of order $O(T)$ contributes. 

Let us now consider the isothermal aging process.
During isothermal aging, there are domain walls separating different
pure states, which are parallel or anti-parallel to a ground state,
at a typical distance $R(\tw)$ from each other. As far as one is
monitoring short time thermal fluctuations due to small scale
droplet excitations, such domain walls appear effectively frozen.
The presence of such frozen-in domain wall influences the small scale
droplet excitations as the following. Some droplets which touch it can
reduce their excitation gap compared with the others in the bulk of
domains. FH considered the probability for such a circumstance to
occur and the amount of reduction of the free energy gap of the
droplet in such a circumstance. As a result, they found that in the
presence of domain wall at a typical distance $R$ from each other the
free energy gap of a droplet of size $L$ becomes on average and
typically as, 
\begin{equation}
    F^{\rm typ}_{L,R} = \Upeff [L/R](L/L_0)^\theta,
\label{eq:domain-1}
\end{equation}
with the effective stiffness constant given by,
\begin{equation}
    \Upeff [L/R] = \Upsilon \left( 1 - c_\upsilon\left({L \over
      R}\right)^{d-\theta}\right), 
\label{eq:Upeff}
\end{equation}
with $c_\upsilon$ being a numerical constant. 
So as the system ages, it appears more and more {\it stiff}.

Now we consider the behavior of the spin auto-correlation function in
the presence of the domain walls.  It can be obtained simply by
replacing the typical free-energy excitation gap $F_L^{\rm typ}$
given in eq.(\ref{eq:FL-FH}) by the reduced one $F^{\rm typ}_{L,R}$
given in eq.(\ref{eq:domain-1}).
Particularly, we assume that the probability distribution of the
free-energy gap $F_{L,R}(i)$ in the presence of the domain walls,
which we denote as $\rho_{L,R}(F)$, follows the same scaling form as
$\rho_{L}(F)$ described in eq.(\ref{eq:rho-FH}),
\begin{equation}
       \rho_{L,R}(F) = {1 \over F^{\rm typ}_{L,R}}
       {\tilde \rho}\left( {F \over  F^{\rm typ}_{L,R}} \right).  
\label{eq:rho-FH-mod}
\end{equation}
Replacing the distribution of the gap $\rho_{L}(F)$ in 
eq.(\ref{eq:cal-Cor-2}) by $\rho_{L,R}(F)$ defined above, we can
obtain the modified behavior of the spin auto-correlation function
in the quasi-equilibrium regime $L(\tau) \ll R(\tw)$ as,
\begin{eqnarray}
    1 & - & C(\tau; \tw) \nonumber \\
  & \sim & 
  T \int_{L_0}^{\infty} {{\rm d}L \over L} \Phi \left(\frac{L}{L(\tau)} \right)
     \rho_{L,R}(0)\\
\label{eq:cal-Cor-3}
& \sim &  1 - C_{\rm eq}(\tau) + 
{\tilde \rho}(0) \frac{T}{\Upsilon(L(\tau)/L_{0})^{\theta}} \nonumber \\
& \times &\int_{L_0/L(\tau)}^{\infty} \frac{dy}{y}  \frac{\Phi(y)}{y^{\theta}}
\left(\frac{\Upsilon}{{\Upsilon}^{\rm eff}[y L(\tau)/R(\tw)]} 
-1 \right) \nonumber \\
   & \sim & 1 - C_{\rm eq}(\tau) + 
\frac{c_1 \tilde{\rho}(0) T}{\Upsilon(L(\tau)/L_0)^\theta}
\left( \frac{L(\tau)}{R(\tw)} \right)^{d-\theta}+\ldots,\nonumber \\ 
\label{eq:cal-Cor-5}
\end{eqnarray}
where the numerical constant $c_{1}$ is given by,
\begin{equation}
c_1=c_{v}\int_{0}^{\infty}dy y^{2(\frac{d-1}{2}-\theta)} \Phi(y).
\label{eq:c_1}
\end{equation}
In  eq.(\ref{eq:cal-Cor-5}), we expanded the effective stiffness given 
in eq.(\ref{eq:Upeff}) assuming that $L(\tau)/R(\tw)$ is small enough
in the quasi-equilibrium regime we are considering here.
The last term is the major correction within the quasi-equilibrium
regime to the equilibrium behavior described by the first two terms.
The lower limit of the integration $L_{0}/L(\tau)$ is put to $0$
assuming $L(\tau)$ is larger enough than $L_0$. Note that the 
integrand in eq.(\ref{eq:c_1}) vanishes as $y \rightarrow 0$ because 
of the inequality $(d-1)/2 > \theta$ \cite{FH-88-EQ}. 
The integral is finite as far as the function $\Phi(x)$ defined in
eq.(\ref{eq:filter}) decays fast enough for large $x$.

The second term  $C_{\rm eq}(\tau)$ in eq.(\ref{eq:cal-Cor-5}) is the 
correlation function in equilibrium obtained by FH \cite{FH-88-EQ},
i.e., $C(\tau)$ in eq.(\ref{eq:cal-Cor-2}),
\begin{equation}
     C_{\rm eq}(\tau) = \lim_{\tw \rightarrow \infty}C(\tau; \tw)
\sim q_{\rm EA} + c_2 \frac{T{\tilde \rho}(0)}
{\Upsilon(L(\tau)/L_0)^{\theta}},
\label{eq:Ceq}
\end{equation}
with $q_{\rm EA}$ being the EA order parameter given by
\begin{equation}
  q_{\rm EA} \equiv  \overline{ \langle S_i \rangle_T^2 } 
 \simeq 1- c_0 \frac{T{\tilde \rho}(0)}{\Upsilon}, 
\label{eq:qEA}
\end{equation}
where $c_i$'s are numerical constants. 

Let us next discuss briefly the relation between the correlation
function $C(t,t')$ with $t=\tau+\tw$ and $t'=\tw$ above described and
magnetic responses, or susceptibilities, frequently measured in
experiments. In the quasi-equilibrium regime, even with weak violation
of the time-translational invariance, the fluctuation-dissipation
theorem (FDT) in the following form is expected to hold at least 
approximately between $C(t,t')$ and the response function $G(t,t')$,
\begin{equation}
G(t,t') = {1 \over T}{\partial C(t,t') \over \partial t'}.
\label{eq:FDT-1}
\end{equation}
From this equation the out-of-phase component of the ac susceptibility
$\chi''(\omega; \tw)$ is evaluated as 
\begin{eqnarray}
     \chi''(\omega; \tw) & = &\int_{\tw - \pi/\omega}^{\tw + \pi/\omega}
     {\rm d}t{\rm sin}(\omega t) \int_0^t{\rm d}t'G(t,t'){\rm
       cos}(\omega t') \nonumber \\
     & \simeq & {\omega \over 2T}{\hat C}(\omega; \tw),
\label{eq:FDT-2}
\end{eqnarray}
where ${\hat C}(\omega; \tw)$ is the Fourier component of 
$C(\tau; \tw)$. The latter is estimated, to a good approximation, 
as~\cite{Fisher-87}  
\begin{equation}
  {\hat C}(\omega; \tw) \simeq - \left.{\pi \over |\omega|} 
{d \over d{\rm ln}\tau} C(\tau; \tw) \right|_{\tau=\tau_\omega},
\label{eq:CeqOmega}
\end{equation}
with $\tau_\omega = 2\pi/\omega$. Hence we obtain
\begin{equation}
     \chi''(\omega; \tw) \simeq - \left.{\pi \over 2T}
    {\partial \over \partial {\rm ln} \tau}
     C(\tau;\tw)\right|_{\tau = \tau_\omega}.
\label{eq:chi-C-1}
\end{equation}
In the above derivation we have used the fact that $C(\tau; \tw)$, as
a function of $\tau$ and $\tw$, varies quite slowly in a time scale of
$\tau_\omega$. The fact that the FDT of eq.(\ref{eq:FDT-2}) holds well 
in the quasi-equilibrium regime, i.e., for 
$R(\tw) \gg L(\tau_\omega)$, has already been tested by the careful
experimental study of the spontaneous magnetic fluctuation which was
directly compared with the relaxation of the ac 
susceptibility.~\cite{AlbaHOR-87,RefregierOHV-88}
The latter experiments have concluded that while there are obvious 
$\tw$-dependences on both the magnetic fluctuations and ac
susceptibility within the quasi-equilibrium regime,
eq.(\ref{eq:FDT-2}) is satisfied in the regime $\omega \tw \gg  1$
within the accuracy of the experiments.

Substituting eqs.(\ref{eq:cal-Cor-4}) and (\ref{eq:Ceq})
into eq.(\ref{eq:chi-C-1}), we obtain,
\begin{equation}
  {\chi''(\omega; \tw) - \chi''_{\rm eq}(\omega) \over 
\chi''_{\rm eq}(\omega) } 
\propto\left( {L(\tau_\omega) \over R(\tw)} \right)^{d-\theta},
\label{eq:resl-chi}
\end{equation}
where $\chi''_{\rm eq}(\omega)$ is the equilibrium ac susceptibility
given by 
\begin{equation}
  \chi''_{\rm eq}(\omega)  \sim  { {\tilde \rho}(0) \over \Upsilon }
    \left (\frac{L_0}{L(\tau_\omega)}\right)^{\theta}
 \frac{d L(\tau)/d{\rm ln} \tau |_{\tau = \tau_\omega}}{L(\tau_\omega)}.
  \label{eq:resl-chieq}
\end{equation}
The above equation can be put into a simple form suggested by 
FH~\cite{FH-88-NE} using eq.(\ref{eq:Upeff});
\begin{equation}
{\chi''(\omega; \tw) - \chi''_{\rm eq}(\omega) \over 
\chi''_{\rm eq}(\omega) } ={\Upeff[L(\tau_\omega)/R(\tw)]^{-1} -
\Upsilon^{-1} \over \Upsilon^{-1} }.
\end{equation}
This formula has been used in the experimental analysis of the
relaxation of the ac susceptibility in a 2D SG 
system.~\cite{Schins-93} 

Let us also note that from the FDT eq.(\ref{eq:FDT-1}) we obtain the
relation between $C(t,t')$ and the zero-field cooled susceptibility 
$\chi_{\rm ZFC} (\tau;\tw)$ as
\begin{equation}
\chi_{\rm ZFC} (\tau; \tw) = \int_{\tw}^{\tau + \tw}{\rm d}t'G(\tau+\tw, t') 
= {1 \over T}[1 - C(\tau; \tw)]. 
\label{eq:FDT-chi}
\end{equation}
Thus $1 - C(\tau; \tw)$, which we will analyze in detail in the next
section, is directly related to $\chi_{\rm ZFC} (\tau;\tw)$ in the
quasi-equilibrium regime. In the aging regime, on the other hand, the
FDT no longer holds. The relation between $G(t,t')$ and $C(t,t')$ in
this regime has been extensively studied, mostly from the view-point
of the mean-field theory,~\cite{CK} but this problem is beyond a scope
of the present work.

To summarize, correction to the ideal-equilibrium behavior of the spin 
auto-correlation function is obtained based on the concept of the
effective stiffness proposed by FH.
The correction is a function of $L(\tau)/R(\tw)$ and
remains small in the quasi-equilibrium regime $L(\tau)/R(\tw) \ll 1$.
Here $L(\tau)$ is the typical size of droplet excitations which can
flip within the time period $\tau$ and $R(\tw)$ is the typical
separation between the domain walls at waiting time $\tw$.
The time dependence of $L(\tau)$ is given by eq.(\ref{eq:L_tau-FH}).
In the droplet theory by FH,~\cite{FH-88-NE} the growth law of 
$R(\tw)$ is assumed to be the same (see eq.(\ref{eq:Rt-FH}) ) based on
the expectation that domain growth in spin glasses are driven by
successive droplet excitations. On the time scale such that 
$L(\tau) \sim R(\tw)$, the above concept which separates the frozen-in 
domain walls and droplet excitations should break down, and the system
crossovers to the aging regime at $R(t=\tau+\tw) \gg R(\tw)$ where
much stronger non-stationarity is expected.

Let us finish the review on the scaling theory with the following
remarks. The above scaling form of $C(\tau; \tw)$ written in terms of 
$R(\tw)$ and $L(\tau)$ is expected to hold more generally than the 
case described above.~\cite{FH-88-EQ} Particularly, we expect it to 
hold for a certain class of the functional form of $L(\tau)$ (and 
so $R(\tw)$), which is determined by the probability distribution 
function $\Psi_L(B)$ in the above argument: a different $\Psi_L(B)$ 
from eq.(\ref{eq:Psi-FH}) associates with a different $L(\tau)$ 
from eq.(\ref{eq:L_tau-FH}) which still ends up with the same scaling 
form of $C(\tau; \tw)$ written in terms of $L(\tau)$ and $R(\tw)$.
Actually, in our previous work I we found that $\Psi_L(B)$ extracted 
from distribution of the largest relaxation time of small systems 
with sizes $L$ has width which does not grow with $L$ in 
contradiction to the scaling form of eq.(\ref{eq:Psi-FH}). However, 
this discrepancy is not crucial in deriving the scaling form of 
$C(\tau; \tw)$. In fact, from the restriction $\tau_{L}(i) < \tau$ in
eq.(\ref{eq:cal-Cor-1}) alone, one finds the same expression for 
$1-C(\tau;\tw)$ as eq.(\ref{eq:cal-Cor-3}) with $L(\tau)$ given by 
eq.(\ref{eq:Ltau-sim}). 

\section{Results of Simulations}

\subsection{Model and method}

We have carried out MC simulations on isothermal aging 
phenomena in the same 3D Ising SG model as in our previous 
work I, i.e., the one with Gaussian nearest-neighbor interactions 
with zero mean and variance $J=1$.
The heat-bath MC method we use here is also the same as in I. 
The SG transition temperature is numerically
determined most recently as $\Tc=0.95\pm 0.04$.~\cite{Marinari-cd98-PS}
The data we will discuss below are obtained at $T=0.5 \sim 0.8$ in 
systems with linear size $L_{\rm s}=24$ averaged over 160 samples with 
one MC run for each sample. 
In the previous work I, it was confirmed that finite size effects do
not appear for these parameters within our time window ($\nle 2 \times 
10^{5}$ MCS).

\subsection{Spin auto-correlation function}

\begin{figure}
\leavevmode\epsfxsize=80mm
\epsfbox{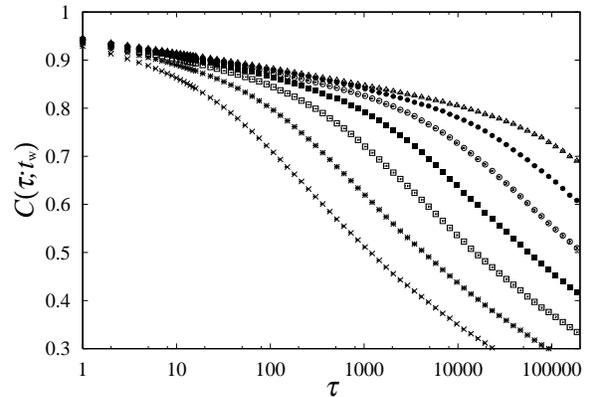}
\caption{The spin auto-correlation function $C(\tau; \tw)$  at $T=0.6$ 
as a function of $\tau$ for fixed $\tw =2^{(2n+3)}$ with $n=1 \ (\tw=32), 
2, ...,7 (\tw=131072)$ from bottom to top, simulated on systems with
$L_{\rm s}=24$. 
}
\label{fig:CisoT6}
\end{figure}

In Fig.~\ref{fig:CisoT6}, we show an overall profile of the
raw data of the spin auto-correlation function $C(\tau; \tw)$ 
obtained by the present MC simulation. 
As can be seen in the figure, the curves of different waiting time 
$\tw$ exhibit a characteristic crossover depending on $\tw$ as 
already found in the previous 
studies.~\cite{Rieger-95,Kisker-96,Rieger-93} The data contain both 
the quasi-equilibrium regime and aging regime, and the crossover 
between the two regimes is expected to occur at around $\tau \sim \tw$. 

According to the droplet theory, $1-C(\tau; \tw)$ in the 
quasi-equilibrium regime is given by eq.(\ref{eq:cal-Cor-5}) with  
a contribution of the equilibrium part $1-C_{\rm eq}(\tau)$ and 
a correction term which is proportional to
$(L(\tau)/R(\tw))^{d-\theta}$.  Using eq.(\ref{eq:Ltau-sim}) for 
$L(\tau)$ in eq.(\ref{eq:Ceq}), the equilibrium part is expected to 
behave as,
\begin{equation}
C_{\rm eq}(\tau) \propto \qea + c (\tau/\tau_{0})^{-\theta/z(T)}.
\label{eq:Ceq=qea+power}
\end{equation}
On the other hand, for the aging regime 
$R(t=\tau+\tw) \gg R(\tw)$ the droplet theory~\cite{FH-88-NE}
predicts that $C(\tau; \tw) \propto (R(\tw)/R(t=\tau+\tw))^\lambda$ 
where $\lambda$ is a new dynamical exponent which satisfies  
$\lambda \ge d/2$.
Rieger~\cite{Rieger-93} has pointed out that with eq.(\ref{eq:Rt-sim}) 
for $R(t)$,  one obtains,
\begin{equation}
C(\tau; \tw) \propto (t/\tw)^{-{\tilde \lambda}(T)},
\label{eq:rieger}
\end{equation} 
with ${\tilde \lambda}(T)=\lambda/z(T)$.
This interpretation has allowed one to explain the apparent temperature
dependence of the exponent ${\tilde \lambda}(T)$ of the data in the
aging regime obtained by MC simulations on 3D Ising SG
models.~\cite{Rieger-93,Kisker-96}
For instance, using the values of exponent ${\tilde \lambda}(T)$
reported in Fig.2 of ref.~\cite{Rieger-93} and our eq.(\ref{eq:z_T}),
$\lambda$ is evaluated at $T < 0.7$ as $1.5 \sim 1.6$ which is rather 
close to its lower bound $d/2=3/2=1.5$. 

\begin{figure}
\leavevmode\epsfxsize=80mm
\epsfbox{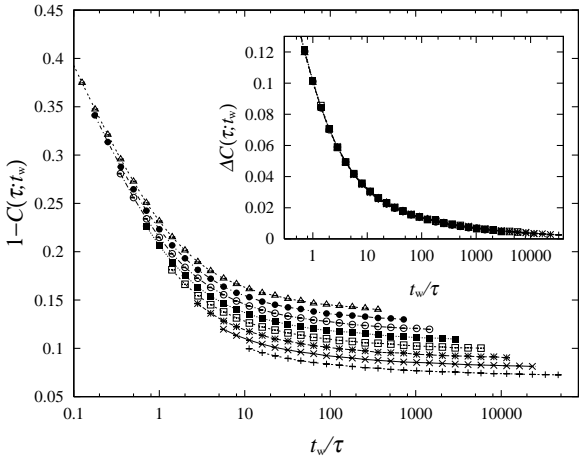}
\caption{The function $\DelCone{}$ at $T=0.6$ 
as a function of $\tw/\tau$ for fixed $\tau = 2^{(2+n)}$ with 
$ n=0,1,...,7$ from bottom to top. The values of $\tw$, for which we 
have simulated $C(\tau; \tw)$, are $\tw = 2^{(10+n)/2}$ with 
$n=1 \ (\tw=45), 2, ..., 25 (\tw=185360)$. 
When these data with each $\tau$ are vertically shifted by the 
amount $\alpha(\tau)$ properly chosen, they lie top on each other 
as shown in the inset where $\DelC{} = \DelCone - \alpha(\tau)$.
}
\label{fig:Iso2-f2}
\end{figure}

Our interest in the present work is the details of $C(\tau; \tw)$ in 
the quasi-equilibrium regime. In order to investigate the latter 
regime, it is convenient to look at the spin auto-correlation 
function $C(\tau; \tw)$ as a function of $\tw$ with $\tau$ considered
as a parameter. In Fig.~\ref{fig:Iso2-f2} $\DelCone{}$ 
of some fixed small $\tau$'s  are plotted against $\tw/\tau$. Note
that in this figure data at larger $\tw/\tau$ are closer to the
equilibrium value as in the same sense in the experimental
measurements of the ac susceptibilities during isothermal aging.
According to the above criterion, the quasi-equilibrium 
regime is the part $\tw/\tau \gg 1$ where the droplet theory suggests
scaling forms such as eqs.(\ref{eq:cal-Cor-5}), (\ref{eq:resl-chi})
and (\ref{eq:resl-chieq}).

\subsection{Scaling analysis}

The scaling ansatz eq.(\ref{eq:cal-Cor-5}) suggests that the data 
of $\DelCone{}$ can be decomposed into the equilibrium part, which does 
not depend on the waiting time $\tw$ and the correction term
which depends on both $\tau$ and $\tw$.  Let us denote the former as 
$\alpha(\tau)$ and the latter as $\DelC{}$, i.e., 
\begin{equation}
\DelCone{} \equiv \DelC{}+\alpha(\tau),
\label{eq:def-delc}
\end{equation}
with
\begin{equation}
\alpha(\tau) \equiv 1-C_{\rm eq}(\tau),
\label{eq:alpha}
\end{equation}
and
\begin{equation}
\DelC{}\equiv c_1 \tilde{\rho}(0)
\frac{T}{\Upsilon}\left ( \frac{L_{0}}{L(\tau)}\right)^\theta
\left( \frac{L(\tau)}{R(\tw)} \right)^{d-\theta}.
\label{eq:delc}
\end{equation}

The correction term $\DelC{}$ contains the factor 
$(L_{0}/L(\tau))^{\theta}$. For the data we have simulated
this factor is proportional to $(\tau/\tau_{0})^{-\theta/z(T)}$. 
The temperature dependence of exponent $1/z(T)$ is given in 
eq.(\ref{eq:z_T}). The energy exponent was obtained in our previous work 
I as $\theta = 0.20 \pm 0.03$.
Therefore the exponent $\theta/z(T)$ becomes quite small, and so the
ratio of the above factor of the maximum time separation $\tau$ and to 
that of the minimum $\tau$ of the data shown in Fig.~\ref{fig:Iso2-f2} is
close to unity.  Actually with $\tau_{\rm max}/\tau_{\rm min}=2^{7}$
and $T=0.6$, the ratio is estimated as $0.94$. This observation
indicates that the factor $(L_{0}/L(\tau))^{\theta}$ can be
approximated as 
\begin{equation}
\left( {L_0 \over L_T(\tau)} \right)^\theta 
= \left( {\tau_0 \over \tau} \right)^{\theta/z(T)}
\simeq c\left[1 - { \theta \over z(T) }{\rm ln}\left({\tau \over 
\tau_{\rm m}} \right)\right], 
\label{eq:theta/z}
\end{equation}
where $\tau_0$ is a characteristic time scale corresponding to 
the length scale $L_0$, $\tau_{\rm m} \simeq 
(\tau_{\rm min}\tau_{\rm max})^{1/2}$, and $c$ a numerical constant 
($\simeq (\tau_0/\tau_{\rm m})^{\theta/z(T)}$). In eq.(\ref{eq:delc}), 
in particular, it can be treated practically as a constant.

Because of the above simplification, the correction term
$\DelC{}$ practically becomes a function of the scaled length 
$L(\tau)/R(\tw)=(\tau/\tw)^{1/z(T)}$ alone. We now try to 
determine $\alpha(\tau)$ for each $\tau$ so that a master curve is 
obtained when $\DelC{}$ of different $\tau$ are plotted against the
rescaled time $\tw/\tau$. Postponing the description of details of the 
analysis till after eq.(\ref{eq:def-kappa}) below, we demonstrate a 
typical result in the inset of Fig.~\ref{fig:Iso2-f2}. As can be seen
in the figure, this one-parameter scaling does work very well.  
\begin{figure}
\leavevmode\epsfxsize=80mm
\epsfbox{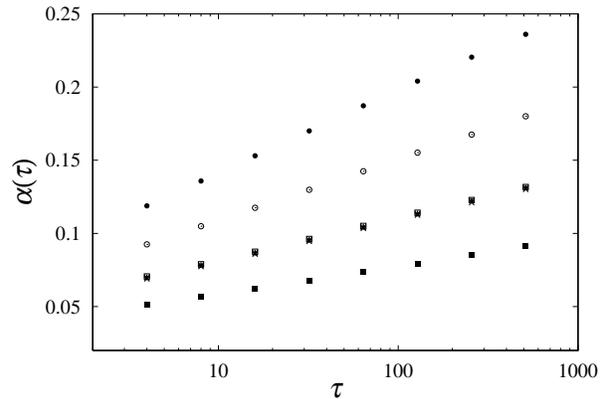}
\caption{The extracted $\alphat$ of eq.(\ref{eq:alpha}) at 
$T=0.5, 0.6, 0.7$ and $0.8$ from bottom to top. For $T=0.6$ four 
sets of $\{ \alphat \}$ are drawn (see the text after 
eq.(\ref{eq:def-kappa})).
}
\label{fig:alph-iso}
\end{figure}

The chosen $\alphat$ are drawn in Fig.~\ref{fig:alph-iso}, where we 
present also the result of the same analysis on $C(\tau; \tw)$ at 
$T=0.5, 0.7$ and $0.8$ with $1/z(T)=0.085, 0.115$ and $0.127$, 
respectively, as obtained in I. They appear as linear functions of 
ln$\tau$.  This nearly logarithmic increase of $\alphat$ can be 
understood as the following. Combining eqs.(\ref{eq:Ceq}) and 
(\ref{eq:qEA}) in \S 2, 
we obtain
\begin{eqnarray}
\alphat & = &1-C_{\rm eq}(\tau)  \sim   
\frac{T{\tilde \rho}(0)}{\Upsilon}
\left[ 1 - \left(\frac{L_0}{L(\tau)}\right)^{\theta} \right] \nonumber \\
& \sim & \frac{cT{\tilde \rho}(0)}{\Upsilon}\frac{\theta}{z(T)}
\ln \left(\frac{\tau}{\tau_{m}} \right) + {\rm const.},
\label{eq:alph-cal}
\end{eqnarray}
where we used eq.(\ref{eq:theta/z}) to derive the last expression. 
The above formula implies that slopes of $\alpha(\tau)$ with respect
to $\ln(\tau)$ depend on temperature as $T^{2}$ because of the 
temperature dependence of the exponent $1/z(T)$ as given in 
eq.(\ref{eq:z_T}). From the data at four temperatures we obtained 
that $\partial \alpha(\tau)/\partial {\rm ln}\tau \propto T^{2.4}$. 
The stronger $T$-dependence than $T^2$ may be attributed to that of 
${\tilde \rho}(0)/\Upsilon$ in the last expression of 
eq.(\ref{eq:alph-cal}). 

It is noted that the second expression of eq.(\ref{eq:alph-cal}) 
implies that $1-C_{\rm eq}(\tau)$ is concave in ln$\tau$ as far as 
the EA order parameter $\qea$ is finite. But such a tendency cannot be 
seen in time-window of the present analysis. No tendency of saturation 
to $\qea$ is seen also in Fig.~\ref{fig:CisoT6} where an overall 
profile of the spin auto-correlation function obtained by the present
simulation up to $\tau \nle \tw \sim 2 \times 10^5$ is shown. In fact 
up to now, no numerical data of the spin auto-correlation function 
which exhibit any tendency of saturation to some $\qea$ have been 
reported for the 3D SG models in spite of the enormous 
computational efforts~\cite{Rieger-93,Ogielski}. This is another
pre-asymptotic behavior of the simulated data.

\begin{figure}
\leavevmode\epsfxsize=80mm
\epsfbox{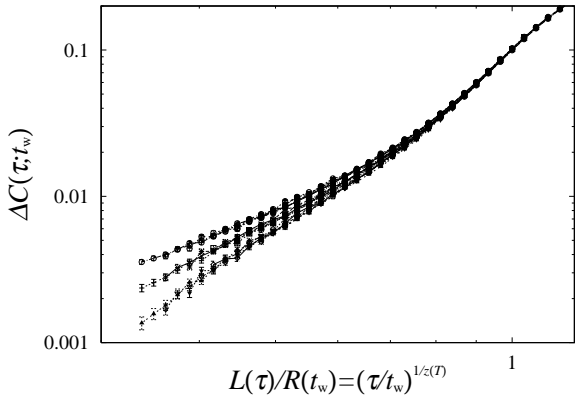}
\caption{The double logarithmic plots of $\DelC{}$ at $T=0.6$ versus 
$L(\tau)/R(\tw)$ with three different sets of $\{ \alphat \}$:
$\alpha_1 \equiv \alpha(\tau=4)=0.0690, 0.0702$ and $0.0712$ from top
to bottom, which yield $\kappa \simeq 2.2, 3.0$ and $3.3$,
respectively.  
}
\label{fig:mcorIr3A}
\end{figure}

Lastly let us discuss the exponent $\kappa$ defined by, 
\begin{equation}
\DelC{} \sim (L(\tau)/R(\tw))^\kappa,
\label{eq:def-kappa}
\end{equation}
at the limit $L(\tau)/R(\tw) \ll 1$.  The question is if it
agrees with the expected behavior given in eq.(\ref{eq:delc}), 
i.e., $\kappa=d-\theta$. 
It is noted for clarity that the limitation of the computational 
time window does not allow us to take equilibrium limit 
$L(\tau)/R(\tw) \rightarrow 0$ accurately and thus leaves some 
ambiguities in the determination of $\kappa$. 
In Fig.~\ref{fig:mcorIr3A}, for examples, we show the double 
logarithmic plots of $\DelC{}$ at $T=0.6$ versus 
$L(\tau)/R(\tw)=(\tau/\tw)^{1/z(T)}$ for three different sets of 
$\{ \alphat \}$. Here we used $1/z(T=0.6) \simeq 0.102 $ using
eq.(\ref{eq:z_T}). 
\begin{figure}
\leavevmode\epsfxsize=80mm
\epsfbox{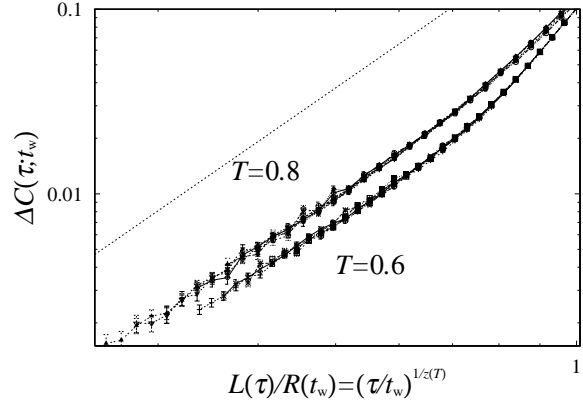}
\caption{The double logarithmic plot of $\DelC{}$ versus 
$L(\tau)/R(\tw)$. The slope of the dotted line is  $d\ (=3)$.
}
\label{fig:mcIre386}
\end{figure}

To draw these figures we have first chosen the value $\alpha(\tau=4) 
\ (=\alpha_1)$ and then determined $\alphat$ of the larger $\tau$ 
subsequently in such a way that the data of $\DelC{}$ lie on a single master
curve. When we chose a too small value for $\alpha_1$, such as the case with
$\alpha_1=0.0690$ shown in the figure, we obtain a small $\kappa \ 
(\simeq 2.2)$ but a linear portion of the double logarithmic plot 
almost disappears. A too large $\alpha_1 \ (=0.0712)$, on the other 
hand, ends up with a large $\kappa\ (\simeq 3.3)$
but the plot deviates from a linear behavior at smaller
$(L(\tau)/R(\tw))$. With a median value of $\alpha_1\ (=0.0702)$, which
yields $\kappa \simeq 3.0$, we obtain linear behavior of the plot in a 
reasonable range of $L(\tau)/R(\tw)$. Thus the acceptable values 
for $\kappa$ from our simulated data are $\kappa \simeq 
2.3 \sim 3.1$. The latter values are consistent with the 
expected exponent in eq.(\ref{eq:delc}), $d-\theta$ with 
$\theta \simeq 0.20$. 
However, it is hard to extract the value of $\theta$ 
by the data of $\DelC{}$ alone. Let us remark however
that the above ambiguity in $\kappa$ little affects the values of 
$\{ \alphat \}$, in particular, their slopes with respect to 
ln$\tau$ are not sensitive. 
This is demonstrated in Fig.~\ref{fig:alph-iso} by 
drawing four sets of $\{ \alphat \}$ at $T=0.6$ with 
$\alpha_1=0.0690, 0.0698, 0.0702$ 
and $0.0712$ which yield $\kappa \simeq 2.2, 2.6, 3.0,$ and $3.3$, 
respectively.

These circumstances are similar at other temperatures we have examined. 
Therefore we show the representative results of $\{ \alphat \}$ in 
Fig.~\ref{fig:alph-iso} which give rise to $\kappa \simeq 2.8 \sim 
3.0$. In Fig.~\ref{fig:mcIre386} we show $\DelC{}$ at $T=0.6, 0.8$ 
with $\kappa \simeq 3.0$. For $\DelC{}$ with a common $\kappa$ but
with different temperatures $T_1$ and $T_2$, their relative magnitude
in the range $L(\tau)/R(\tw) \ll 1$ can be estimated by that of
$(\tilde{\rho}(0)T/\Upsilon)_{T=T_i}$ through eq.(\ref{eq:delc}). The
relative magnitude of the two $\DelC{}$ shown in
Fig.~\ref{fig:mcIre386} is about $1.3$ which is roughly consistent
with this estimation ($ = (0.8/0.6)^{1.4}\simeq 1.50$).

\subsection{Out-of-phase linear ac susceptibility}

\begin{figure}
\leavevmode\epsfxsize=80mm
\epsfbox{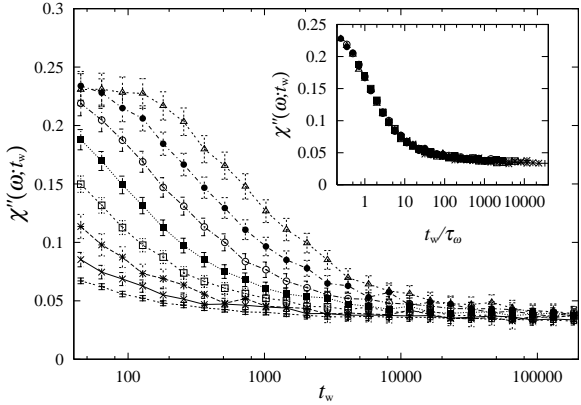}
\caption{$\chiomegatw$ evaluated through eq.(\ref{eq:chi-C-1}) 
for $\tau_\omega = 2^{(2+n)}$ with $ n=0,1,...,7$ from bottom to
top. In the inset the same data are plotted against $\tw/\tau_\omega$.
They lie on a universal curve.
}
\label{fig:chi-org}
\end{figure}

Let us finally discuss briefly the logarithmic derivative of the
auto-correlation function, which is supposed to be almost 
identical to the out-of-phase linear ac susceptibility via FDT.
Here let us denote the logarithmic derivative of the
auto-correlation function simply as out-of-phase 
linear ac susceptibility $\chi''(\omega; \tw)$.
We show in Fig.~\ref{fig:chi-org} 
$\chiomegatw$  at different $\omega$ which are 
obtained by numerical differentiation of eq.(\ref{eq:chi-C-1}) using 
the same data of $1-C(\tau; \tw)$ shown in Fig.~\ref{fig:Iso2-f2}.  

Using the growth laws eqs.(\ref{eq:Rt-sim}) and (\ref{eq:Ltau-sim}) 
in eqs.(\ref{eq:resl-chi}) and (\ref{eq:resl-chieq}) respectively, 
the expected scaling behavior is obtained as,
\begin{equation}
  {\chi''(\omega; \tw) - \chi''_{\rm eq}(\omega) \over 
\chi''_{\rm eq}(\omega) } 
\propto \left(\frac{\tau_\omega}{\tw} \right)^{(d-\theta)/z(T)},
\label{eq:dchi-power}
\end{equation}
with 
\begin{equation}
  \chi''_{\rm eq}(\omega) 
\sim (\tau_{0}\omega)^{\theta/z(T)}\chi''_{\rm eq}(\omega_{0}),
\label{eq:chi-power}
\end{equation}
where $\omega_{0}=2\pi/\tau_{0}$.
Note that the exponent $\theta/z(T)$ in eq.(\ref{eq:chi-power}) 
originates from the factor $(L_0/L(\tau))^{\theta}$ in 
eq.(\ref{eq:resl-chieq}), and that the range of time 
$\tau_\omega=2\pi/\omega$ of interest here is the same as that in 
the analysis of $1-C(\tau;\tw)$. Because of the same reason as we 
noted above eq.(\ref{eq:theta/z}), a master curve for 
$\chi''(\omega;\tw)$ has been obtained by simply plotting them 
against $\tw/\tau_\omega$ as shown in  the inset of 
Fig.~\ref{fig:chi-org}.  

Let us remark that $\chiomegatw$ simulated here are apparently 
larger for the larger $\tau_\omega$ (smaller $\omega$). 
The scaling ansatz suggests that this is not the case for $\chiomegatw$
of much smaller frequencies, whose equilibrium values 
$\chi''_{\rm eq}(\omega)$ do depend on the frequencies and are the 
smaller for the smaller frequencies, in their asymptotic approach to 
the equilibrium values. We also note that, in order to check for the 
FDT to hold in the quasi-equilibrium regime it is desirable to 
perform direct simulation on the response against ac magnetic field. 
Since such an analysis is time consuming, it is left for a future 
work.

\section{Discussions}

We have shown so far that the scaling properties of the simulated 
auto-correlation function $C(\tau; \tw)$ in the quasi-equilibrium 
range $L(\tau) \ll R(\tw)$ can be well explained by the scaling 
ansatz by the droplet theory expressed in terms of $L(\tau)$ and 
$R(\tw)$. For the growth law of the latter, we have used the 
power-law form, eqs.(\ref{eq:Rt-sim}) and (\ref{eq:Ltau-sim}) for 
$R(\tw)$ and $L(\tau)$, respectively. More precisely, the fundamental
exponents $\theta$ and $\kappa\ (=d-\theta)$ in the droplet theory,
combined with the exponent $1/z(T)$ of the growth-law of $R(\tw)$,
consistently describe our simulated results: the energy  relaxation in
I and aging behavior of $C(\tau; \tw)$ in the quasi-equilibrium
regime in the present work. However, the time window of the present 
simulation and the previous one in I are in the range where the growth 
law of $R(\tw)$ exhibits the pre-asymptotic behavior expressed as
feature 2) in \S 1.
Time scales of laboratory experiments on real spin glasses may also be
in a similar range as we will discuss below. 

The quantity, which has been studied by experiments most frequently 
and is of interest from our present point of view, is the ac 
susceptibility $\chiomegat$. The latter is measured continuously in 
time $t$ while the system is isothermally aged. Since the period 
$\tau_\omega = 2\pi/\omega$ is, in general, much smaller than the 
aging time $t$, it should be regarded as the response in the 
quasi-equilibrium range ($t=\tw+\tau_\omega \simeq \tw$) as we noted 
before. Indeed the scaling analysis based on eq.(\ref{eq:resl-chi}) 
was directly carried out by Schins {\it et al} on a 2D Ising 
SG system.~\cite{Schins-93} Their results are consistent 
with the 2D version of the droplet theory\cite{FH-87-2D} and, quite 
remarkably, the asymptotic growth law eq.(\ref{eq:Rt-FH}) with a 
reasonable value of $\psi$.

Concerning the the ac susceptibility $\chiomegat$ in the 3D spin 
glasses, the Saclay group has often reported that $\chiomegat$ obey 
the $\omega t$-scaling; if $\chiomegat$ with different $\omega$ are 
vertically shifted by certain amount depending on $\omega$, and are 
plotted against $\omega t$, they lie on a universal 
curve.~\cite{VincHO,VincBHL} The equilibrium values 
$\chi''_{\rm eq}(\omega)$ thus obtained obey a power-law
scaling,~\cite{VincHO,AlbaHOR-87,Saclay-90} 
\begin{equation} 
\chi''_{\rm eq}(\omega) \propto \omega^{\hat{\alpha}}.
\label{eq:alpha-Vin}
\end{equation}
The $\omega t$-scaling for $\chiomegat$ can be naturally explained 
by the scaling ansatz eq.(\ref{eq:resl-chi}) with the power-law form 
for $R(t)$ and $L(\tau_\omega)$, i.e., eq.(\ref{eq:dchi-power}). 
However, the data reported are not enough for examining the exponent 
$\hat{\kappa}\equiv \kappa/z(T) =(d-\theta)/z(T)$. 
The above equation (\ref{eq:alpha-Vin}) just corresponds to our 
eq.(\ref{eq:chi-power}) with $\hat{\alpha}=\theta/z(T)$. The time 
window of the experiments are $\tw \nle 10^{16}$ and $10^{10} \nle 
\tau_\omega \nle 10^{14}$ in unit of  microscopic time ($\sim \Tc$), 
while that of our simulation are $\tw \nle 10^5$ and $4 \le 
\tau_\omega \le 256$. Presumably this difference allows one to 
extract the exponent $\hat{\alpha}$ from the experimentally measured
$\chiomegat$.  
Interestingly, ${\hat \alpha}$ of CdIn$_{0.3}$Cr$_{1.7}$S$_4$ 
reported in Fig. 2b of Ref.~\cite{Saclay-90} is rather in good 
agreement with $\theta/z(T)$ evaluated by our simulational studies 
on both the order of magnitude and the temperature dependence, though 
this is not the case at temperatures close to $\Tc$ nor at low 
temperatures $T \nle 0.5\Tc$.
In contrast to the above experimental results which are favorable to 
the power-law form of $R(\tw)$ and $L(\tau)$, the Saclay group also 
reported that $\chi''_{\rm eq}(\omega)$ observed in a wider range 
($7$ decades) of $\omega$ fits rather well to a function of ln$\omega$ 
than to the power-law form of eq.(\ref{eq:chi-power}).~\cite{Saclay-90}

Svedlindh {\it et al},~\cite{Svedlindh-92} reported a 
ln$t$-dependence of $\chiomegat$ and a power-law dependence on 
$\omega$ in the logarithmic derivative of $\chiomegat$ as, 
\begin{equation}
 { \partial \chiomegat \over \partial {\rm ln}t } 
  \sim \omega^{-\hat{\beta}}, 
 \label{eq:beta-Sve}
\end{equation}
at $T<\Tc$ ($\hat{\beta} \simeq 0.25 $ at $T=0.985 \Tc$). 
This feature may also be explained by eq.(\ref{eq:resl-chi}) so long 
as $L(\tau_\omega)$ obeys the power-law form eq.(\ref{eq:Ltau-sim}). 
The exponent  $\hat{\beta}$ in eq.(\ref{eq:beta-Sve}) is then 
evaluated as $\hat{\beta} = (d-2\theta)/z$, which is $[(d/\theta)-2]$
times larger than $\hat{\alpha}$ in eq.(\ref{eq:chi-power}). 

From our point of view, another important quantity is the 
zero-field-cooled susceptibility $\chi_{\rm ZFC}(\tau; \tw)$ or 
the thermoremanent magnetization $m_{\rm TRM}(\tau; \tw)$ in 
the quasi-equilibrium regime. The two quantities are related by 
$\chi_{\rm ZFC}(\tau; \tw) = [m_{\rm FC} - 
m_{\rm TRM}(\tau; \tw)]/h$, where $m_{\rm FC}$ is the field-cooled 
magnetization under a field of strength $h$, and 
$\chi_{\rm ZFC}(\tau; \tw)$ is directly related to 
$1 - C(\tau; \tw)$ as expressed by eq.(\ref{eq:FDT-chi}) in \S 2. 
Although $\chi_{\rm ZFC}(\tau; \tw)$ (or $m_{\rm TRM}(\tau; \tw)$) has 
been been measured frequently since the pioneering 
work by Lundgren {\it et al},~\cite{Lundgren} most of the 
measurements have been focused on its crossover behavior between 
the quasi-equilibrium and aging regimes. Recently we have analyzed 
the data on $m_{\rm TRM}(\tau; \tw)$ of AgMn spin glass measured 
by the Saclay group~\cite{VincBHL} by the same procedure as we did 
on $1 - C(\tau; \tw)$ in \S 3. The results are qualitatively 
similar to those of $\chiomegat$ described just above: for the 
data at temperatures not close to $\Tc$, the orders of magnitude 
of $\hat{\kappa}$ and $\hat{\alpha}$ as well as their temperature 
dependence are compatible with those obtained by our simulations, 
though this is not the case for the data at temperatures close to 
$\Tc$. 

Besides the experiments on the linear ac susceptibilities, 
Joh et al~\cite{Joh-98} have recently extracted $R(t)$ from the 
thermoremanent magnetization they have measured under fields of 
various strengths. Its growth-law agrees with the simulational 
result of eq.(\ref{eq:Rt-sim}) even quantitatively, and the 
estimated $R(t)$ with $t$ of the order of laboratory time is 
at most only several tens of lattice distances, i.e., a length 
scale far from the thermodynamic limit we usually imagine. 

Thus concerning the growth law of $R(t)$ and $L(\tau)$ in 3D spin 
glasses, the experimental studies so far reported have not converged 
to a unique conclusion: it is either a power-law as 
eq.(\ref{eq:Rt-sim}) or a logarithmic form as eq.(\ref{eq:Rt-FH}). 
This ambiguous situation seems difficult to be resolved by looking 
at only $\omega$ dependence of $\chi''_{\rm eq}(\omega)$ because 
$\hat{\alpha}$ in eq.(\ref{eq:alpha-Vin}) is quite small partially 
because of the smallness of $\theta$. The latter had already made 
it difficult to settle serious controversies concerning static 
properties of the SG phase.~\cite{BM-85-com}
In this respect, it is of interest to examine experimentally the  
$\omega t$-dependence of $\chi''(\omega;t)$ more in detail because
$\hat{\beta}$ in eq.(\ref{eq:beta-Sve}) is relatively large. Also, 
as mentioned above, $\chi_{\rm ZFC}(\tau; \tw)$ in the 
quasi-equilibrium regime may be another important quantity to be 
studied in detail. 
Such experiments have to be performed in a time window of
observations as wide as possible. 

Among problems on isothermal aging phenomena to be further explored 
numerically, a most important one may be aging dynamics at 
temperatures close to $\Tc$, where a crossover from aging dynamics 
governed by the criticality at $\Tc$ to that governed by the $T=0$ 
criticality (the droplet theory) is expected to 
occur.~\cite{FH-88-EQ} Indeed such a crossover in aging dynamics 
has been found recently in the 4D Ising SG model by Hukushima 
{\it et al}.~\cite{Huku} A similar analysis on the present 3D model 
may shed light on peculiar behavior of aging dynamics measured also 
experimentally at temperatures close to $\Tc$ as we have pointed out
above.  

To conclude, we have simulated isothermal aging in the 3D Ising 
SG model, and analyzed behaviors of the spin auto-correlation 
function in the quasi-equilibrium regime. The simulated results 
are well interpreted by the scaling arguments which associate with the 
characteristic length scales $R(\tw)$ and $L(\tau)$, i.e, those of
domains and droplet excitations in time scales of $\tw$ and $\tau$, 
respectively. We have conjectured that some aspects of the experimental
data on the ac susceptibility and the zero-field-cooled susceptibility 
can be also explained along the same line.

\section*{Acknowledgments}
We would like to E. Vincent and M. Ocio for their fruitful discussions 
and for kindly sending us their data on AgMn spin glass. We also thank
J.-P. Bouchaud, K. Hukushima and P. Nordblad for their useful 
discussions. Two of the present authors (T. K. and H. Y.) were supported 
by Fellowships of Japan Society for the Promotion of Science for 
Japanese Junior Scientists.
This work is supported by a Grant-in-Aid for International Scientific
Research Program, ``Statistical Physics of Fluctuations in Glassy
Systems'' (\#10044064) 
and by a Grant-in-Aid for Scientific Research Program (\#10640362),
from the Ministry of Education, Science and Culture.
The present simulation has been performed on FACOM VPP-500/40 at the 
Supercomputer Center, Institute for Solid State Physics, the University of 
Tokyo.

\end{document}